\documentclass[showpacs,amssymb,aps,twocolumn]{revtex4}
\usepackage{amsmath}
\usepackage{amstext}
\usepackage{amsopn}
\usepackage{amsfonts}
\usepackage{amssymb}
\usepackage{bbm}
\usepackage{accents}
\usepackage{empheq}
\usepackage{graphicx}
\usepackage{epsf}
\usepackage{graphics}
\usepackage[latin1]{inputenc}
\def\sc{\scriptscriptstyle}

\newcommand{\be}{\begin{equation}}
\newcommand{\ee}{\end{equation}}
\newcommand{\bea}{\begin{eqnarray}}
\newcommand{\eea}{\end{eqnarray}}

\begin{document}

\title{Phenomenological implications of $S$-duality symmetry}

\author{Ashok Das$^{a,b}$ and Jnanadeva  Maharana$^{c,d}$\footnote{$\ $ e-mail: das@pas.rochester.edu,  maharana@iopb.res.in}}

\affiliation{$^a$ Department of Physics and Astronomy, University of Rochester, Rochester, 
NY 14627-0171, USA}
\affiliation{$^b$ Saha Institute of Nuclear Physics, 1/AF Bidhannagar, Calcutta 700064, India}
\affiliation{$^{c}$ Institute of Physics, Bhubaneswar 751005, India}
\affiliation{$^{d}$ National Institute of Science Education and Research, Bhubaneswar 751005, India}

\date{\today}

\begin{abstract}
It is proposed that $S$-duality is a fundamental symmetry of nature which is spontaneously broken. 
Axion and dilaton are identified with the doublet of the $S$-duality symmetry 
group $SL(2,\mathbbm{R})$. The symmetry is broken at a high scale corresponding to the 
experimentally estimated axion decay constant $f_{\chi}$. The symmetry breaking
mechanism is discussed in analogy with PCAC in pion physics. 
$S$-duality invariant interactions of fermions with axion and dilaton doublet are  
introduced. The symmetry breaking mechanism contributes negligibly small
corrections to fermion masses in the QCD sector. Inspired by universality in string theory,
the $S$-duality invariant interaction of the axion-dilaton doublet to QCD fermions is proposed to generalize  
to all fermions. Phenomenological consequences of this broken symmetry are explored.
\end{abstract}

\maketitle

\section{Introduction}

The standard model of particle physics describes the three fundamental forces 
of nature (excluding gravitation) in the microscopic domain. The electro-weak 
 theory explains the weak as well as the electromagnetic phenomena 
\cite{g,s,w,gt,v} and quantum chromodynamics (QCD) is the accepted theory of 
strong interactions \cite{gr,p,wl}. The standard model has been tested to a  
great degree of accuracy and there are no serious discrepancies between the 
theoretical predictions and the experimental results at energies below the 
weak scale.  However, it is believed that the standard model is incomplete 
in the sense that it contains several arbitrary parameters and does not 
incorporate gravitational interactions. The grand unified theories (GUTs) 
were proposed to provide a unified description of the strong and the 
electro-weak phenomena \cite{ps,gg,ms1,ms2}; however, proton decay, which 
is a crucial prediction of all GUTs, has not been observed as yet. Furthermore,
 while string theory is a promising candidate to unify all the four 
fundamental interactions (including gravity), the standard model of particle 
physics in totality is yet to emerge from string theory. 

With the running of LHC now, there is great excitement in probing energy 
regimes beyond the weak scale. This will allow us to observe certain massive 
particles which are an essential integral part of the standard model and to 
study their properties. For example, the Higgs boson which is responsible for 
the spontaneous breakdown of the electro-weak symmetry is expected to be 
discovered at LHC. Supersymmetry which, among other things, leads to a 
resolution of the hierarchy problem is expected to be tested at LHC as well. 
There are, of course, a host of other models which have been proposed as 
describing physics beyond the standard model and experiments at LHC will 
likely decide on their validity. One of the characteristic (and bothersome) 
features of all endeavors to construct a model beyond the standard model has 
been the proliferation of particle spectra at the very fundamental level. 
Guided mainly by symmetry considerations we propose here a model with only 
one extra particle which may provide a window to access physics beyond the 
standard model. Some of our predictions can possibly be tested in precision 
experiments at LHC, underground experiments as well as in cosmological 
experiments.

To introduce our model, let us recall that strong interactions had a heavy 
mass puzzle commonly known as the $U_{\sc A} (1)$ problem \cite{wein}. 
The resolution of this problem within the context of QCD was shown by 't Hooft 
\cite{gt1} to be related to the existence of instanton solutions which 
essentially change 
the structure of the QCD vacuum. The instanton solutions effectively lead to 
an additional term in the QCD Lagrangian density so that we have
\begin{equation}
{\cal L} = - {\rm Tr} \left(\frac{1}{2g^{2}}\, F_{\mu\nu}F^{\mu\nu} + 
\frac{\theta g^{2}}{16\pi^{2}}\,F_{\mu\nu}\widetilde{F}^{\mu\nu}\right),
\label{yang-mills}
\end{equation}
where the field strength tensor is a matrix in the fundamental representation 
of $SU(3)$ and we have scaled out the Yang-Mills coupling constant $g$ from 
the field strength tensor. The dual of the field strength tensor is defined 
to be
\begin{equation}
\widetilde{F}^{\mu\nu} = \frac{1}{2}\,\epsilon^{\mu\nu\lambda\rho} 
F_{\lambda\rho},\label{dual}
\end{equation}
and $\theta$ (in \eqref{yang-mills}) represents the parameter commonly known as the QCD vacuum angle. 
However, in the complete standard model including the electro-weak theory, 
the phases of the quark mass matrix $M$ also contribute to the pseudoscalar 
density in \eqref{yang-mills} and this is proportional to Arg $\det M$. 
Therefore, the parameter $\theta$ in \eqref{yang-mills} is to be understood 
as $\theta_{\rm eff} = \theta + {\rm Arg} \det M$. The $\theta$ dependent 
term in \eqref{yang-mills} is a total divergence and, therefore, does not 
modify the equations of motion for the gauge fields. However, it violates $P$ 
and $T$ (and, consequently, $CP$) and induces an electric dipole moment for 
the neutron. The very stringent limits on the neutron electric dipole 
moment measurements \cite{dipole}  
limit the value of the effective $\theta$ parameter (we continue to write 
$\theta$ for simplicity) to $\theta < 10^{-9}-10^{-10}$. Without any 
natural (theoretical) justification for why (effective) $\theta$ should be 
so small, the resolution of the $U_{\sc A}(1)$ problem leads to the strong 
$CP$ problem.

A possible resolution of the strong $CP$ problem which is also quite 
attractive from the cosmological point of view is through the Peccei-Quinn 
mechanism \cite{pq}. Here one promotes the $\theta$ parameter to a dynamical 
pseudoscalar field, the axion $\chi$, in a way such that the theory has 
an additional global $U(1)$ chiral invariance. This symmetry is spontaneously 
broken (through a choice of the axion potential) so that the axion develops a 
vacuum expectation value (vev) leading to the $\theta$ parameter. Being the 
vev of a field, this can always be adjusted to be small which is the rationale 
for the solution of the strong $CP$ problem through the Peccei-Quinn mechanism.
The axion is the Nambu-Goldstone boson \cite{weinb, wilcz}
of the broken $U(1)$ symmetry introduced by Peccei and Quinn. 
Although the original model of Peccei-Quinn has been experimentally ruled 
out, generalizations of the model where the axion is a weakly interacting, 
long lived and a very light mass particle (also known as the invisible axion) 
remain quite attractive. We refer the readers to some of the review articles 
\cite{r1,r2,r3,r4,ms,rdp,kim,db,rdp1,sik}
for details and for the literature in this subject. The present bounds on the 
axion parameters are given by
\begin{equation}
m_{\chi} \simeq 10^{-4} {\rm eV},\quad f_{\chi} \simeq 10^{9}-10^{12} 
{\rm GeV},\label{axmass}
\end{equation}
where $f_{\chi}$ denotes the axion decay constant (into two gluons). The coupling of the axion 
to the pseudoscalar  density in such models has the generic form
\begin{equation}
{\cal L}_{\chi} = - \frac{\zeta}{f_{\chi}}\, 
\frac{g^{2}}{16\pi^{2}}\,{\rm Tr} \left(\chi F_{\mu\nu}
\widetilde{F}^{\mu\nu}\right),\label{axionL}
\end{equation}
where $\zeta$ denotes a model dependent constant parameter and the vev of the 
axion is related to $\theta$ through a multiplicative factor. A direct search for axions, however, 
has not been successful thus far. We would like to 
propose a model based on $S$-duality symmetry which can be tested at LHC and 
if experimental observations validate the symmetry it may lead to an 
indirect proof of the axion.

Let us note that the Lagrangian density \eqref{yang-mills} can be rewritten 
in the form
\begin{align}
{\cal L} & = - \frac{1}{16 \pi}\, {\rm Im}\bigg({\rm Tr} 
\Big( \tau (F^{\mu\nu}\pm i{\widetilde F}^{\mu\nu})
(F_{\mu\nu}\pm i{\widetilde F}_{\mu\nu})\Big)\bigg)\notag\\
 & = - \frac{1}{16 \pi}\, {\rm Im}\bigg({\rm Tr} \Big( \pm i\tau 
F^{\mu\nu\pm} \widetilde{F}^{\pm}_{\mu\nu}\Big)\bigg),\label{ym}
\end{align}
where $\tau= \pm \frac{\theta g^2}{2\pi}+\frac{4 i\pi}{g^2}$ denotes the moduli 
parameter and $F_{\mu\nu}^{\pm} = F_{\mu\nu} \pm i \widetilde{F}_{\mu\nu}$. 
Note that the angle $\theta$ has a period of $2\pi$ and 
the resulting equations of motion are invariant under the 
transformations 
$\tau\rightarrow \tau+1$ as well as $\tau\rightarrow -\frac{1}{\tau}$. In 
fact, under the fractional transformations $SL(2,\mathbbm{Z})$  
\begin{equation}
\tau\rightarrow \frac{a\tau+b}{c\tau +d},\label{fractional}
\end{equation}
which can be thought of as a discrete $S$-duality transformation, the field 
strength tensors transform as
\begin{equation}
\tau F^{+}_{\mu\nu}\rightarrow (a\tau +b)F^{+}_{\mu\nu},~~
{\bar\tau}F^{-}_{\mu\nu}\rightarrow(a{\bar\tau}+b)F^{-}_{\mu\nu},\label{sdual}
\end{equation}
where $a,b,c,d\in\mathbbm{Z}$ with $ad-bc=1$ and a ``bar" (over $\tau$) denotes complex 
conjugation. 
The action associated with (\ref{ym}) is not invariant under \eqref{fractional}-\eqref{sdual}, 
but the equations of motion are. 
 
It should be emphasized that this symmetry in the equations of motion 
manifests only in the Yang-Mills sector and it will be gratifying if we can 
promote this to the fully interacting theory including fermions (even if 
at the level of equations of motion). We note that such $S$-duality symmetries  
appear naturally in supergravity as well as in string theories  
(more details can be found in the review articles \cite{sugra}). 
In the present context we plan to generalize the discrete transformation in 
\eqref{fractional}-\eqref{sdual} to a continuous global $S$-duality transformation by introducing 
only one additional scalar field which is the $S$-duality partner of the axion. 
This partner scalar field couples to the scalar Yang-Mills Lagrangian density 
(much like the axion couples to the pseudoscalar density in \eqref{axionL}) 
in such a way that the theory is invariant under the continuous $S$-duality 
transformations. We call this $S$-duality partner (of axion), a scalar field 
$\phi$, with the {\it dilaton}. We emphasize here that we do not intend 
to invoke any string theoretic argument for the origin of this scalar field. In 
fact, a scalar field with a very light mass has also been introduced in the 
context of Jordan-Brans-Dicke theory of gravity \cite{j,bd}. 
It is adequate,  for the model we 
propose, that this should be a light, weakly interacting scalar field which is the $S$-duality partner of the axion, independent of any other identification. In fact, 
being the partner of the axion $\chi$, the $\phi$ field is expected to share 
most of the 
attributes of the axion, it will be a weakly interacting, long lived and a 
very light mass particle. Since this global symmetry is not observed 
in nature,
 it will be broken (leaving possibly a discrete symmetry at the level of 
equations of motion) and we envision breaking this symmetry along the 
PCAC scenario much as in pion physics \cite{lee,geff}.

We remind the reader that $S$-duality has been a very fertile
domain of research in quantum field theory, supersymmetric theories,
supergravity and string theories. There are very robust results, especially in
theories with a large number of supersymmetries and in supergravity theories.
Furthermore, in the context of string theory, $S$-duality has  played a
cardinal role in the investigation of nonperturbative properties of
string theories. Moreover, a host of very important results in stringy black 
hole
physics are derived by exploiting this symmetry. We believe that $S$-duality 
is a fundamental symmetry of nature although it is broken. Our goal is more 
pragmatic and our approach is more phenomenological although we are clearly 
inspired by the powers of $S$-duality symmetry in various fields. We propose a 
model based on $S$-duality symmetry and hope that some of the consequences 
of our
proposal can be tested in experiments at LHC as well as in cosmological
experiments. 

The plan of the paper is as follows. In section {\bf II}, we describe the 
$S$-duality invariant Lagrangian densities for the axion-dilaton system as 
well as their 
coupling to fermions. In section {\bf III}, we discuss the mechanism 
for the breaking of the $S$-duality symmetry. It is along the lines of PCAC in 
the study of pions. We derive the consequences following from such a scenario. 
In Section {\bf IV}, we argue that the $S$-duality invariant coupling to 
fermions is universal for all fermion species including the neutrinos. This is 
in the spirit of string theory where the dilaton couples univerally to all 
excitation of the string (its vev is the coupling constant of string theory) 
and axion, being its $S$-duality partner, couples universally as well. In this 
scenario neutrinos can acquire a small mass due to the
spontaneous breaking of $S$-duality symmetry (independent of whether the 
neutrino is a Dirac or a Majorana particle).

\section{$S$-duality and axion-dilaton actions}

In this section, we intend to present our model in some detail. We will 
follow a more efficient approach and construct $S$-duality invariant actions. 
We choose   $SL(2,\mathbbm{R})$ as  our $S$-duality group. (We note here that the group $SL (2,\mathbbm{R})$ is isomorphic to the symplectic group $Sp(2, \mathbbm{R})$ as well as to the generalized unitary group $SU (1,1)$.) The Lie algebra  $s\ell(2)$ (which is isomorphic to the Lie algebra  $su (1,1)$) is given by 
\bea
\label{sualgebra}
[T_1,T_2]=-iT_3,\quad [T_2,T_3]=iT_1,\quad 
[T_3,T_1]=iT_2,
\eea
which differs from the $su (2)$ Lie algebra in the sign of the first 
commutator and is connected with the fact that $SL(2,\mathbbm{R})$ is a noncompact 
group (unlike $SU (2)$). As a result,  it is not possible to have a finite 
dimensional unitary representation of $SL (2,\mathbbm{R})$ and we cannot choose all the 
finite dimensional generators $T_{1}, T_{2}, T_{3}$ of the group to be Hermitian. For our 
purposes, it is sufficient to look at two dimensional representations of 
the group and a choice for the generators can be taken to be
\begin{align}\label{sumatrices}
& T_1= \frac{i}{2} \sigma_{1} = \frac{1}{2}
\begin{pmatrix}0 & i\\ i & 0 \end{pmatrix},\quad T_2
= -\frac{i}{2} \sigma_{3} = \frac{1}{2}
\begin{pmatrix}-i & 0\\ 0 & i \end{pmatrix},\nonumber\\
&  T_3= \frac{1}{2} \sigma_{2} =  \frac{1}{2} 
\begin{pmatrix}0 & -i\\ i & 0 \end{pmatrix},\quad T_{i}^{\dagger}T_{3} = 
T_{3} T_{i},
\end{align}
where $i=1,2,3$ and  $\sigma_{1}, \sigma_{2}, \sigma_{3}$ represent the three 
Pauli matrices. An alternative choice for the generators can be taken to be
\begin{align}\label{sumatrices1}
& \overline{T}_1= \frac{i}{2} \sigma_{1} = 
\frac{1}{2}\begin{pmatrix}0 & i\\ i & 0 \end{pmatrix},\quad \overline{T}_2
= \frac{i}{2} \sigma_{2} = \frac{1}{2}
\begin{pmatrix}0 & 1\\ -1 & 0 \end{pmatrix},\nonumber\\
&  \overline{T}_3= \frac{1}{2} \sigma_{3} =  \frac{1}{2} 
\begin{pmatrix}1 & 0\\ 0 & -1 \end{pmatrix},
\end{align}
However, all such choices correspond to a change of basis and are related by 
a unitary transformation. For example, the choices of the generators in 
\eqref{sumatrices} and \eqref{sumatrices1} are related by the unitary 
matrix ($SS^{\dagger} = \mathbbm{1} = S^{\dagger}S$)
\begin{equation}
S = \frac{1}{\sqrt{2}} \left(\mathbbm{1} - i \sigma_{1}\right) = 
\frac{1}{\sqrt{2}} \begin{pmatrix}
1 & -i\\
-i & 1
\end{pmatrix},\ ST_{i}S^{\dagger} = \overline{T}_{i},\label{smatr}
\end{equation}
with $i=1,2,3$.

A $2\times 2$ matrix representing a general $SL (2,\mathbbm{R})$ transformation has the 
form (in a given basis)
\begin{equation}
\Omega = e^{-i \alpha_{k} T_{k}},\label{su11tfn}
\end{equation}
where $\alpha_{k}, k = 1,2,3$ denote the three real constant (global) parameters of the 
transformation. Note that the transformation matrix in \eqref{su11tfn} is real (since the generators are purely imaginary), namely,
\begin{equation}
\Omega^{*} = \Omega,\quad \Omega^{\dagger} = \Omega^{T}.\label{reality}
\end{equation}
Since a finite dimensional representation of $SL (2,\mathbbm{R})$ 
is not unitary, the transformation matrices satisfy the condition 
(see \eqref{sumatrices})
\begin{equation}
\Omega^{\dagger} T_{3}\Omega = \Omega^{T} T_{3} \Omega = T_{3},\quad \Omega^{T} T_{3} = 
T_{3} \Omega^{-1},\label{grmetric}
\end{equation}
so that $(2T_{3})$ can be thought of as the metric in the group space 
(note that $(2T_{3})^{2} = \mathbbm{1}$). Under a finite $SL (2,\mathbbm{R})$ 
transformation, a vector (in this case a doublet) transforms as
\begin{equation}
\Psi \rightarrow \Omega \Psi,\label{doublettfn}
\end{equation}
while a matrix in the adjoint representation transforms as
\begin{equation}
M \rightarrow \Omega M \Omega^{T}.\label{adjointtfn}
\end{equation}

Let us choose the dilaton and the axion fields parameterizing the coset 
$\frac{SL (2,\mathbbm{R})}{U(1)}$ in the form
\begin{align}
V & = \begin{pmatrix} 
e^{-\phi} + \chi^{2} e^{\phi} & & \chi e^{\phi}\\
\chi e^{\phi} & & e^{\phi}
\end{pmatrix} = \begin{pmatrix}
v_{0} + v_{2} & v_{1}\\
v_{1} & v_{0} - v_{2}
\end{pmatrix}\nonumber\\
& = v_{0} \mathbbm{1} - 2i v_{1} T_{1} + 2i v_{2} T_{2},\label{vmatrix}
\end{align}
which would transform under a $SL (2,\mathbbm{R})$ transformation in the adjoint 
representation as in \eqref{adjointtfn}
\begin{equation}
V \rightarrow \Omega V \Omega^{T}.\label{vtfn}
\end{equation}
Here we have identified
\begin{align}
v_{0} & = \frac{1}{2}\big(e^{-\phi} + \chi^{2} e^{\phi} + e^{\phi}\big),
\nonumber\\
v_{1} & = \chi e^{\phi},\nonumber\\
v_{2} & = \frac{1}{2}\big(e^{-\phi} + \chi^{2} e^{\phi} - 
e^{\phi}\big).\label{vcomponents}
\end{align}
It follows from \eqref{vcomponents} that
\begin{equation}
v_{0}^{2} - v_{1}^{2} - v_{2}^{2} = 1,\label{constraint}
\end{equation}
which is a reflection of the condition 
\begin{equation}
\det V = 1,
\end{equation}
which holds for a special linear matrix. The matrix $V$ in \eqref{vmatrix} 
is easily seen to satisfy  (see \eqref{grmetric})
\begin{equation}
V^{T} = V,\quad VT_{3} = T_{3} V^{-1}.
\end{equation}
We note here that the dilaton and the axion fields (as well as the matrix $V$) 
introduced in \eqref{vmatrix} have zero canonical dimension. However, we can 
relate them to conventional spin zero fields with unit canonical dimension in 
$3+1$ dimensions (see, for example, \eqref{axionL}) through the simple rescaling 
\begin{equation}
\phi \rightarrow \frac{\phi}{f_{s}},\quad \chi \rightarrow 
\frac{\chi}{f_{s}},\label{scaling}
\end{equation}
where $f_{s}$ denotes the scale of breaking for the  $S$-duality symmetry 
(we will use this later).

As we have already mentioned in the introduction, the action describing the coupling of the dilaton 
and the axion to the Yang-Mills field cannot be written in a manifestly 
Lorentz and $S$-duality invariant manner although the equations of motion are. 
In this context following comments are in order. One can write a manifestly 
$S$-duality ($SL (2,\mathbbm{R})$) invariant Lagrangian density at the expense of 
manifest Lorentz invariance. Furthermore, it is necessary to introduce 
auxiliary gauge fields for this purpose. However, when these auxiliary fields 
are eliminated, the dynamical equations reduce to manifestly $S$-duality 
invariant and Lorentz invariant equations as noted in the introduction. We 
refer the interested reader to \cite{senjohn,sugra} for more details on this.

The Lagrangian density for the dilaton and the axion can be written in the 
standard manner as 
 \begin{align}
 {\cal L}_{(\chi\phi)} & = - \frac{f_{s}^{2}}{4}\, {\rm Tr}\, \partial_{\mu} V^{-1} \partial^{\mu} V
 \nonumber\\
 & = - \frac{f_{s}^{2}}{2} \left(\partial_{\mu}v_{0} \partial^{\mu}v_{0} - \partial_{\mu}v_{1} \partial^{\mu}v_{1} - \partial_{\mu}v_{2} \partial^{\mu}v_{2}\right) \nonumber\\
 & = \frac{f_{s}^{2}}{2}\left(\partial_{\mu}\phi \partial^{\mu}\phi + e^{2\phi}\, \partial_{\mu}\chi \partial^{\mu}\chi\right),\label{dilaton-axion}
 \end{align}
which is invariant under the $SL (2,\mathbbm{R})$ transformations \eqref{vtfn}. Note 
that under the scaling \eqref{scaling}, the free dilaton part of the 
Lagrangian density in \eqref{dilaton-axion} takes the conventional form of 
a free spin zero boson theory. In terms of the complex moduli
 \begin{equation}
 \tau = \chi + i e^{-\phi},\label{moduli}
 \end{equation}
 this Lagrangian density can also be written as
 \begin{equation}
 {\cal L}_{(\chi\phi)} = - \frac{f_{s}^{2}}{2}\, 
\frac{\partial_{\mu} \tau \partial^{\mu} \bar{\tau}}{(\tau - \bar{\tau})^{2}},
 \end{equation}
 where the moduli parameters (coupling constants)
 introduced in \eqref{ym} can be related to the 
vacuum expectation value of the moduli defined in \eqref{moduli}. The 
$S$-duality invariance of \eqref{dilaton-axion} leads to the N\"{o}ther 
current matrix 
 \begin{equation}
 J^{\mu}_{(\chi\phi)} = - 2i V^{-1} \partial^{\mu} V,
\label{chiphicurrentmatrix}
 \end{equation}
 with the components given by
 \begin{equation}
 J^{\mu}_{(\chi\phi)\, i} = {\rm Tr}\, T_{i} J^{\mu}_{(\chi\phi)}.
\label{chiphicurrent}
 \end{equation} 
 Explicitly the components take the form
 \begin{equation}
 J^{\mu}_{(\chi\phi)\, 1}  = v_{0}\overleftrightarrow{\partial}^{\mu} v_{1},\ J^{\mu}_{(\chi\phi)\, 2} = - v_{0}\overleftrightarrow{\partial}^{\mu} v_{2},\ J^{\mu}_{(\chi\phi)\, 3} = v_{1}\overleftrightarrow{\partial}^{\mu} v_{2}.
 \end{equation}
 
 To write the free fermion action in a $S$-duality invariant manner, let us 
consider a four component  spinor $\psi$ which may be a Dirac or a Majorana 
spinor. Let us next construct a doublet of $SL (2,\mathbbm{R})$ as
 \begin{equation}
 \Psi = \frac{1}{2\sqrt{2}} \begin{pmatrix}
\big((\mathbbm{1}-\gamma_{5}) + i (\mathbbm{1} + \gamma_{5})\big)\psi\\ 
\big((\mathbbm{1}+\gamma_{5}) + i (\mathbbm{1} - \gamma_{5})\big)\psi
\end{pmatrix},\label{fermion}
\end{equation}
where $\gamma_{5} = i\gamma^{0}\gamma^{1}\gamma^{2}\gamma^{3}$ and $\Psi$ 
transforms under $SL (2,\mathbbm{R})$ as a vector \eqref{doublettfn}, namely,
\begin{equation}
\Psi \rightarrow \Omega \Psi.\label{fermiontfn}
\end{equation}
The Dirac adjoint of this doublet (which transforms inversely under a 
Lorentz transformation) has the form
\begin{equation}
\overline{\Psi} = \frac{1}{2\sqrt{2}}\begin{pmatrix}
\bar{\psi} \big((\mathbbm{1} + \gamma_{5}) - i (\mathbbm{1} - \gamma_{5})\big) & \bar{\psi} \big((\mathbbm{1} - \gamma_{5}) - i (\mathbbm{1} + \gamma_{5})\big)
\end{pmatrix}.
\end{equation}
From the point of view of $S$-duality transformation, however, it is more 
useful to define an alternative adjoint of \eqref{fermion} which transforms 
inversely under both Lorentz as well as $S$-duality transformations as
\begin{equation}
\widetilde{\overline{\Psi}} = \overline{\Psi} (2T_{3}).\label{adjoint}
\end{equation}
In fact, under a $SL (2,\mathbbm{R})$ transformation \eqref{fermiontfn}
\begin{equation}
\widetilde{\overline{\Psi}} \rightarrow \overline{\Psi} \Omega^{T} (2T_{3}) = \overline{\Psi} (2T_{3}) \Omega^{-1} = \widetilde{\overline{\Psi}} \Omega^{-1},\label{adjfermiontfn}
\end{equation}
where we have used \eqref{grmetric}. With these we can now write the free 
fermion Lagrangian density as
\begin{equation}
{\cal L}_{\psi} = i \widetilde{\overline{\Psi}} \gamma_{5} \partial\!\!\!\slash \Psi = i \bar{\psi} \partial\!\!\!\slash \psi,\label{fke}
\end{equation}
which is easily seen using \eqref{fermiontfn} and \eqref{adjfermiontfn} to be 
manifestly $SL (2,\mathbbm{R})$ invariant. The N\"{o}ther current from the fermion 
sector can be derived to have the form
\begin{equation}
J^{\mu}_{(\psi)\, i} = \widetilde{\overline{\Psi}} T_{i} \gamma^{\mu} \Psi,\label{fermioncurrent}
\end{equation}
so that the total $S$-duality current is given by (see \eqref{chiphicurrent} 
and \eqref{fermioncurrent})
\begin{equation}
J^{\mu}_{i} = J^{\mu}_{(\chi\phi)\, i} + J^{\mu}_{(\psi)\, i}.
\end{equation}

We can now introduce the interaction of the dilaton and the axion to the 
QCD fermions in a $SL (2,\mathbbm{R})$ invariant manner as follows. Here we have two possibilities. If we want the interaction to be of nongradient type, we can choose the Lagrangian density to be
\begin{equation}
{\cal L}_{1 \sc Y} = - i g_{1{\sc Y}}\Lambda_{\sc\rm QCD}\, \widetilde{\overline{\Psi}} \gamma_{5} V (2T_{3}) \Psi,\label{nongradient}
\end{equation}
which is manifestly $SL (2,\mathbbm{R})$ invariant. Namely,
\begin{align}
{\cal L}_{1 \sc Y} & \rightarrow - ig_{1{\sc Y}}\Lambda_{\sc\rm QCD}\, \widetilde{\overline{\Psi}} \Omega^{-1} \gamma_{5} \Omega V \Omega^{T} (2T_{3}) \Omega \Psi\nonumber\\
& = - ig_{1{\sc Y}}\Lambda_{\sc\rm QCD}\, \widetilde{\overline{\Psi}} \gamma_{5} V (2T_{3}) \Omega^{-1}\Omega\Psi 
= {\cal L}_{1 \sc Y},\label{su11inv}
\end{align}
where we have used \eqref{vtfn}, \eqref{fermiontfn}, \eqref{adjfermiontfn} as well as \eqref{grmetric}. Here we have chosen the coupling constant $g_{1\sc Y}$ to de dimensionless and as a result the interaction strength involves a mass scale appropriate for the interaction and $\Lambda_{\sc\rm QCD}$ denotes the QCD scale which is the appropriate scale for the interactions since we are considering only the QCD sector here. Explicitly, the interaction Lagrangian density has the form
\begin{align}
{\cal L}_{1 \sc Y} & = - g_{1 \sc Y}\Lambda_{\sc\rm QCD}\, v_{2} \bar{\psi} \psi + ig_{1 \sc Y}\Lambda_{\sc\rm QCD}\, v_{1} \bar{\psi}\gamma_{5} \psi\nonumber\\
& = - \frac{g_{1 \sc Y}\Lambda_{\sc\rm QCD}}{2}\left(e^{-\phi} + \chi^{2} e^{\phi}  - e^{\phi}\right) \bar{\psi} \psi\nonumber\\
&\quad  + ig_{1 \sc Y}\Lambda_{\sc\rm QCD}\, \chi e^{\phi} \bar{\psi} \gamma_{5} \psi,\label{yukawa1}
\end{align}
which is manifestly parity conserving since the dilaton $\phi$ is a scalar while the axion $\chi$ is a pseudoscalar.

On the other hand, if we want the interaction to be of gradient type as in the case of the standard axion, we can consider the Lagrangian density
\begin{equation}
{\cal L}_{2 \sc Y} =  - g_{2 \sc Y}\,\widetilde{\overline{\Psi}} 
(\partial\!\!\!\slash V) (2T_{3}) \Psi 
\label{yukawa}
\end{equation}
which can be shown to be manifestly $SL (2,\mathbbm{R})$ invariant following \eqref{su11inv}. Explicitly, the interaction Lagrangian density has 
the form
\begin{align}
{\cal L}_{2 \sc Y} & = - g_{2 \sc Y}\bar{\psi} (\partial\!\!\!\slash v_{0}) \psi,\nonumber\\
& = - \frac{g_{2 \sc Y}}{2} \left(-e^{-\phi} + \chi^{2} e^{\phi} + e^{\phi}\right) \bar{\psi} (\partial\!\!\!\slash \phi) \psi\nonumber\\
&\quad  - g_{2 \sc Y}\,\chi e^{\phi}\, \bar{\psi} (\partial\!\!\!\slash \chi) \psi, \label{intL}
\end{align}
which is manifestly parity conserving. It is worth noting here that even though this is a gradient coupling, the form of the interaction is quite distinct from that in the case of the standard axion (even after symmetry breaking as we will discuss in the next section). Furthermore, we point out here that, as in PCAC, the gradient coupling \eqref{yukawa} can be generated from the free Lagrangian density for the fermion \eqref{fke} by redefining the fermion fields as
\begin{equation}
\Psi\rightarrow e^{ig_{2\sc Y} V (2T_{3})} \Psi,\quad \widetilde{\overline{\Psi}} \rightarrow \widetilde{\overline{\Psi}} e^{-ig_{2\sc Y} V (2T_{3})}.
\end{equation}
However, the nonderivative interaction \eqref{nongradient} cannot be generated through a field redefinition, even from an invariant mass-like (nonderivative) term, since
\begin{equation}
\widetilde{\overline{\Psi}} \Psi = 0 = \widetilde{\overline{\Psi}} \gamma_{5} \Psi.
\end{equation}

\section{Symmetry breaking}

From the energy scales probed so far, there is no compelling experimental 
evidence in favor 
of an exact $S$-duality symmetry in nature. Therefore, this symmetry
needs to be broken and if axion and dilaton are $S$-duality partners, their 
masses are expected to be of the same order of magnitude. Light weakly 
interacting particles (axions) are yet to be directly   
seen in laboratory and cosmological experiments, although there are stringent 
limits on their masses.  For the scalar field $\phi$ which is of stringy origin or corresponds to the Jordan-Brans-Dicke scalar, 
there are also constraints on its mass which are similar to that of the axion. We 
are encouraged by these observations. Furthermore, since the Lagrangian 
density for the axion and the dilaton \eqref{dilaton-axion} is defined on a 
coset much like the nonlinear sigma model for the pions (in the present 
context, the coset space is $\frac{SL(2,\mathbbm{R})}{U(1)}$), we consider the symmetry 
breaking to be completely parallel to the PCAC mechanism in the 
theory of pions. However, there would be important differences because unlike 
the theory of pions where the coset space $\frac{SU(2)\times SU (2)}{SU(2)}$ 
is compact, here the coset space is noncompact. Furthermore, since we would 
like the axion field to have a nonzero vacuum expectation value (in order to 
have a nontrivial theta parameter in the theory), it is clear from the 
constraint \eqref{constraint} that $v_{0}, v_{1}$ need to have nontrivial 
vacuum expectation value and \eqref{constraint} can in principle determine 
the vev for $v_{2}$ to be nonzero as well.  

To understand the vev structures better, let us note that \eqref{constraint} 
can be thought of as describing an (internal) anti-de Sitter space. Therefore, 
let us parameterize the space with hyperbolic coordinates as
\begin{equation}
v_{0} = \cosh \Sigma,\ v_{1} = \sinh \Sigma \cos \xi,\ v_{2} = \sinh \Sigma \sin\xi,\label{hyperbolic}
\end{equation}
where the constraint \eqref{constraint} is automatically satisfied. Comparing with 
\eqref{vcomponents} we can identify
\begin{align}
e^{-\phi} & = \frac{1}{v_{0} - v_{2}} = \frac{1}{\cosh\Sigma - \sinh\Sigma \sin\xi},\nonumber\\
\chi & = \frac{v_{1}}{v_{0} - v_{2}} = \frac{\sinh\Sigma \cos\xi}{\cosh\Sigma - \sinh\Sigma \sin\xi}.\label{angcomponents}
\end{align}
It is clear from this that we can give vevs to both the dilaton as well as 
the axion consistent with \eqref{constraint} by requiring
\begin{equation}
\langle \Sigma\rangle = \Sigma_{0},\quad \langle \xi\rangle = \xi_{0},\label{vev}
\end{equation}
where $\Sigma_{0},\xi_{0}$ are arbitrary constants and these would translate 
into the vevs 
\begin{align}
\langle e^{\phi}\rangle & = e^{\phi_{0}} = \cosh\Sigma_{0} - \sinh\Sigma_{0}\sin\xi_{0},\nonumber\\
\langle \chi\rangle & = \chi_{0} = e^{-\phi_{0}} \sinh\Sigma_{0}\cos\xi_{0}.\label{vev1}
\end{align}
From \eqref{hyperbolic} we see that this can also be written as
\begin{equation}
\langle v_{0}\rangle = \cosh\Sigma_{0},\, \langle v_{1}\rangle = \sinh\Sigma_{0}\cos\xi_{0},\, \langle v_{2}\rangle = \sinh\Sigma_{0}\sin\xi_{0}.\label{vvevs}
\end{equation}

A direct consequence of the vevs is that the two different Yukawa interactions in \eqref{yukawa1} as well as \eqref{yukawa} will generate trilinear interactions involving the fermions and the axion and the dilaton. In addition, \eqref{yukawa1} will generate masses for the QCD fermions upon shifting.  Shifting \eqref{yukawa1} around the vevs \eqref{vev1} in a consistent manner with the scaling in \eqref{scaling}, namely,
\begin{equation}
\phi\rightarrow \phi_{0} + \frac{\phi}{f_{s}},\quad \chi\rightarrow \chi_{0} + \frac{\chi}{f_{s}},\quad e^{-\phi}\rightarrow e^{-(\phi_{0} + \frac{\phi}{f_{s}})},\label{shift}
\end{equation}
we obtain 
\begin{align}
{\cal L}_{1 \sc Y} & = - m_{f}\bar{\psi} \psi + im \bar{\psi}\gamma_{5}\psi - \frac{g_{\sc Y}\Lambda_{\sc\rm QCD}}{f_{s}}\, (\langle v_{2}\rangle \phi + \langle v_{1}\rangle \chi) \bar{\psi}\psi \nonumber\\
& \quad + \frac{ig_{1 \sc Y}\Lambda_{\sc\rm QCD}}{f_{s}}\, e^{\phi_{0}} (\chi + \chi_{0} \phi) \bar{\psi}\gamma_{5}\psi + \cdots,\label{yukawaexp}
\end{align}
where
\begin{equation}
m_{f} = g_{1 \sc Y}\Lambda_{\sc\rm QCD}\, \langle v_{2}\rangle,\quad m = g_{1 \sc Y}\Lambda_{\sc\rm QCD}\, \langle v_{1}\rangle.\label{fermionmass}
\end{equation}
The higher order terms in the interaction in \eqref{yukawaexp} will be suppressed by inverse powers of $f_{s}$.

On the other hand, a shift of \eqref{intL} leads to 
\begin{align}
{\cal L}_{2 \sc Y} & = - \frac{g_{2\sc Y}}{2 f_{s}} \left(-e^{-\phi_{0}} + \chi_{0}^{2} e^{\phi_{0}} + e^{\phi_{0}}\right) \bar{\psi} (\partial\!\!\!\slash \phi) \psi \nonumber\\
& \quad - \frac{g_{2 \sc Y}}{f_{s}}\,\chi_{0}e^{\phi_{0}}\, \bar{\psi} (\partial\!\!\!\slash \chi) \psi + \cdots,\label{shiftedyukawa}
\end{align}
We note here that the axion-nucleon-nucleon coupling constant has been estimated in the past \cite{kaplan} to be 
\begin{equation}
g_{\chi \sc NN} \sim 10^{-12},\label{axioncoupling0}
\end{equation}
up to a model dependent multiplicative constant of the order of $O(1)$. In order to extract numbers from our model, we propose that the axion-quark-quark coupling constants in our model to satisfy
\begin{equation}
g_{1\sc Y}, g_{2 \sc Y} \sim g_{\chi \sc NN} \sim 10^{-12},\label{axioncoupling}
\end{equation}
up to a multiplicative constant so that the axion remains weakly interacting. With a trilinear coupling of this order of magnitude we note that the values of $m_{f},m$ in \eqref{fermionmass} will depend on the values of the parameters $\Sigma_{0},\xi_{0}$. In particular, we note that since $\chi_{0} \sim \theta$ is expected to be small, it follows from \eqref{vev1} that  $\xi_{0}$ is constrained to be very close to $\frac{\pi}{2}$ in which case \eqref{fermionmass} leads to
\begin{equation}
m \approx 0,\quad m_{f} \approx g_{1\sc Y} \Lambda_{\sc\rm QCD}\sinh\Sigma_{0} \approx 10^{-11} {\rm GeV},\label{masscorrection}
\end{equation}
up to a constant (we have approximated $\Lambda_{\sc\rm QCD} = 220 {\rm MeV} \approx 1 {\rm GeV}$). Here we have used the fact that with $\xi_{0}\approx \frac{\pi}{2}$, it follows from \eqref{vev1} that
\begin{equation}
e^{\phi_{0}} \approx \cosh \Sigma_{0} - \sinh\Sigma_{0},
\end{equation}
which leads to $e^{-\Sigma_{0}} = e^{\phi_{0}}$ and if one were to relate $e^{\phi_{0}} \sim g^{2}$ ($g$ is the QCD coupling), then at the QCD scale, $\Sigma_{0}$ has to be small. As a result, we conclude from \eqref{masscorrection} that the correction to the fermion masses in the QCD sector because of $S$-duality breaking is negligibly small.

It is worth noting here that the trilinear coupling in \eqref{shiftedyukawa} involving the axion has a vector structure unlike the usual axial vector coupling of the axion. (As a result, there will be no $\pi\mbox{-}\chi$ mixing resulting from such an interaction.)  Of course, this would have observational consequences. However, more than that the strength of this interaction is proportional to $\theta$ (because of the factor of $\chi_{0}\sim \theta$) and correspondingly will be highly suppressed. In contrast, the dilaton trilinear interaction in \eqref{shiftedyukawa} will be relatively more dominant which is experimentally interesting.

In this scenario, the dilaton and the axion are Goldstone bosons at this 
stage, just like the pions. As a result, their self interactions involve 
derivative terms and they remain massless even after shifting fields around 
the vevs. To introduce masses for the dilaton and the axion, we introduce a 
term of the form
\begin{equation}
{\cal L}_{m} = - \frac{f_{s}^{2}m_{s}^{2}}{2}\,\frac{v_{0}}{v_{0}-v_{3}} = - \frac{f_{s}^{2}m_{s}^{2}}{2} \left(1 + e^{-2\phi} + \chi^{2}\right).\label{linearL}
\end{equation}
Shifting this around the vevs \eqref{vev1} (see \eqref{shift}) we obtain
\begin{align}
{\cal L}_{m} & = -\frac{f_{s}^{2} m_{s}^{2}}{2}\left[\left(1 + e^{-2\phi_{0}} + \chi_{0}^{2}\right) - \frac{2}{f_{s}}\left(e^{-2\phi_{0}} \phi + \chi_{0} \chi\right)\right.\nonumber\\
&\quad\qquad\qquad \left. + \frac{1}{f_{s}^{2}}\left(2e^{-2\phi_{0}} \phi^{2} + \chi^{2}\right) + \cdots\right],\label{mass}
\end{align}
which leads to masses for the axion and the dilaton of the forms  
\begin{equation}
m_{\chi}^{2} = m_{s}^{2},\quad m_{\phi}^{2} = 2 e^{-2\phi_{0}} m_{s}^{2} = 2 e^{-2\phi_{0}} m_{\chi}^{2}.\label{dilatonaxionmass}
\end{equation}
If one were to identify $e^{\phi_{0}} = g^{2}$, the Yang-Mills coupling, then \eqref{dilatonaxionmass} would suggest that the mass of the dilaton would be relatively larger than that of the axion.

\section{Summary and discussion}

In this work, we have proposed $S$-duality as a fundamental symmetry motivated by the issues in QCD 
which have led in the past to the introduction of the axion. We have identified the group $SL (2,\mathbbm{R})$ as the $S$-dulaity group and the ``scalar'" dilaton has 
been introduced as the $S$-duality partner of the pseudoscalar axion. The 
exponential of the vev of the dialton is related to the QCD gauge coupling 
constant and we have introduced the interaction Lagrangian density for the 
fermions in the QCD sector in a $S$-duality invariant manner. However, in a more general setting such as the string theory, the vev of the dilaton is expected to control all coupling 
constants with the dilaton coupling to all matter. In this spirit, we propose 
universality of $S$-duality symmetry so that all fermions including quarks and leptons couple to the axion and dilaton in a $S$-duality symmetric manner as given in \eqref{yukawa1} and  \eqref{intL}. Of course, the coupling strengths and the interaction scale $\Lambda$ can be different for different species. For leptons, for example, one can identify the interaction scale with the weak scale $\Lambda=\Lambda_{\rm weak}$ (unlike \eqref{nongradient} where the relevant scale is $\Lambda = \Lambda_{\sc\rm QCD}$). 

Although our proposal is clearly influenced by string theoretic ideas, it is worth 
remarking here that our approach has been rather phenomenological so that 
the interactions are not derived from a (unique) fundamental theory. For example, as we have already pointed out, in string theory the vev of the dilaton field is expected to determine all the constants of the (low energy) theory (such as the fine structure constant, Yukawa couplings etc) in addition to  the gravitational constant. However, we do not insist on any such requirement beyond 
the possible relation between $e^{\phi_{0}}$ and the Yang-Mills coupling so 
that our goal is rather modest. String theory admits many more massless scalars and pseudoscalars when compactified to lower dimensions and there have 
been several studies on the phenomenological consequences of these moduli 
in string theory \cite{taylor,axgv}. There have also been attempts to establish 
a connection between the QCD axion and the string theoretic axion although 
the stringy axion models have failed to accomodate all the experimental 
bounds so far \cite{r1,witten}. Moreover,  the implications of 
admitting ultra light axions of string theory in the cosmological context have
been addressed recently \cite{ax}. 
In contrast, in our modest phenomenological approach we have only introduced one additional scalar, namely, the dilaton, to write Lagrangian densities in a $S$-duality invariant manner.

Our interpretation of the universality hypothesis is that there is a single Yukawa coupling for all the fermions belonging to the QCD sector (namely, for all quarks) and a single Yukawa coupling for all
the leptons. We accommodate this proposition qualitatively as follows. As we have mentioned before, one can extract the axion-quark-quark coupling constant (which is of the same order of magnitude for the dilaton coupling) as in \cite{kaplan} (from the estimate
of $g_{\chi\sc NN}$) as discussed in \eqref{axioncoupling}. Similarly, we can estimate the Yukawa coupling in the lepton sector from the estimate in \cite{kaplan}
\begin{equation}
g_{\chi ee} \sim 10^{-15},\label{axionleptoncoupling0}
\end{equation}
by requiring, as in \eqref{axioncoupling}, that ($\tilde{g}_{1\sc Y}, \tilde{g}_{2\sc Y}$ are the analogs of $g_{1\sc Y},g_{2\sc Y}$ in the lepton sector)
\begin{equation}
\tilde{g}_{1\sc Y}, \tilde{g}_{2\sc Y} \sim g_{\chi ee} \sim 10^{-15},\label{axionleptoncoupling}
\end{equation}
up to a multiplicative constant of order unity.

The universality hypothesis leads to some interesting consequences. As we have already pointed out, both the (trilinear) Yukawa interactions in \eqref{shiftedyukawa} and \eqref{yukawaexp} of the fermions with the dilaton and the axion are highly suppressed with our identification \eqref{axioncoupling} which we expect to hold in the lepton sector as well (higher order interactions are suppressed further by inverse powers of $f_{s}$). Moreover, since $S$-duality is broken, the fermions acquire a mass, for example, as given in \eqref{fermionmass} (or the analogous formula in the lepton sector). The masses acquired by diverse families of fermions (due to $S$-duality breaking) would be extremely small since the relevant Yukawa couplings are small. We know that all the fermions (except the neutrino) acquire their masses in the standard model (electroweak theory) through their couplings to the Higgs  and the corrections from the $S$-duality breaking in \eqref{fermionmass} to the  masses of these fermions are negligible (see, for example, \eqref{masscorrection} for the QCD sector). However, since the neutrinos are massless in the standard model, the breaking of $S$-duality symmetry can provide a plausible alternative mechanism to generate a small neutrino mass. In fact, this mechanism can generate a mass for either Dirac or Majorana neutrinos and following \eqref{masscorrection} we can estimate this mass to be of the order of
\begin{equation}
m_{\nu} \sim \tilde{g}_{1 \sc Y} \Lambda_{\rm weak} \sinh\Sigma_{0} \approx 10^{-2} {\rm eV},\label{neutrinomass}
\end{equation}
where we have used \eqref{axionleptoncoupling} as well as have approximated $\Lambda_{\rm weak} \sim 246 {\rm GeV} \approx 1 {\rm TeV}$ (keeping in mind that the relations \eqref{axionleptoncoupling}-\eqref{neutrinomass} hold up to multiplicative constants of order unity). (Note that the corrections to the masses of all charged leptons will be of the same order of 
magnitude as \eqref{neutrinomass}, which is negligible,  due to the universality hypothesis in the lepton sector.) In view of the experimental evidence for neutrino oscillations, the mechanism for generating 
a mass for the neutrino is an active area of study. If neutrino does acquire a mass through 
this mechanism, it may have interesting consequences since it would arise at the $S$-duality breaking scale $f_{s}$ which lies  between the GUT scale and the weak scale. 

It is quite possible that axions and dilatons will be produced in LHC due to
bremsstrahlung although the production cross sections will be quite suppressed
compared to those for the Higgs production and the production of SUSY particles. However,
these particles will travel long distances due to their weak coupling to 
matter. There is also a possibility that the axion and the dilaton produced at LHC will
have a chance to get detected in underground and laboratory experiments
which have been set up for dedicated searches of these particles. Moreover,
although the dilaton proposed by us will have weak interactions, as weak as the axion, 
its interaction with matter will be much stronger than that of the stringy dilaton which is the light scalar partner of graviton in string theory and whose role in the cosmological domain has been a topic of considerable interest in the past \cite{edcop, gv1,gv2}. The cosmological implications of 'our proposed' dilaton will also be of interest to examine carefully.

\section*{Acknowledgments}

We would like to thank Professors A. Melissinos and S. Okubo for valuable comments and 
discussions. We have benefited from numerous discussions with Costas Bachas, 
Pankaj Jain, Varun Sahni and  Gabriele Veneziano. 
One of us (A.D.)  is supported in part  by US DOE Grant number 
DE-FG 02-91ER40685. The other author (J.M.) is primarily supported by
the People of the Republic of India and also partly through a Raja Ramanna 
Fellowship by the Indo-French Center for the Promotion of Advanced Research:
IFCPAR Project No. IFC/4104-2/2010/201.

\end{document}